\documentclass{caosp}

\usepackage{graphicx}

\begin{document}

%
\hauthor{I.M.\,Brand\~{a}o, M.S.\,Cunha and J. F. Gameiro}

\title{Bolometric Correction of the roAp star $\alpha$ Cir}


%
\author{
        I.M.\,Brand\~{a}o \inst{1,} \inst{2},
        M.S.\,Cunha \inst{1}
	\and
	J.F.\,Gameiro \inst{1,} \inst{2}
       }

%
\institute{
           Centro de Astrof\'{i}sica da Universidade do Porto, Rua das estrelas, \\
 4150-762 Portugal (E-mail: isa@astro.up.pt)
         \and 
           Department of Aplied Mathematics, Universidade do Porto, Portugal
          }

\date{}

\maketitle

\begin{abstract}
For the first time, the bolometric correction of $\alpha$ Cir was determined. Two values, both based on an estimation of the total integrated flux, were obtained. For that purpose spectroscopic and photometric data of $\alpha$ Cir available in the literature was used. The values derived were then used to place $\alpha$ Cir in the HR diagram.

\end{abstract}

%
\section{Introduction}
\label{intr}

$\alpha$ Cir (HD 128898) is the prototype of a class of Ap stars that oscillate in high frequencies
(roAp, rapidly oscillating Ap, \cite{kurtz82}). Besides knowing its effective temperature and parallaxe, to accurately place $\alpha$ Cir in the HR diagram it is necessary to know also its bolometric correction ($B.C.$). To determine the latter, the method proposed by \cite{north81} was applied.

\section{Bolometric Correction determination and discussion}

The $B.C.$ is given by the equation,

\begin{equation}
B.C.=-2.5\log\int\limits_0^\infty\,F(\lambda)\mathrm{d}\lambda-m_V-11.492,
\label{eq:BC}
\end{equation} 
where $m_V$ is the apparent visual magnitude and $F(\lambda)$ is the flux at a given wavelength. 
Two values for the integrated flux of $\alpha$ Cir were obtained. The first was determined by combining the observed ultraviolet flux of $\alpha$ Cir retrieved from IUE Newly Extracted Spectra (INES) data archive, with the theoretical flux obtained from the Kurucz model (with IDL routine KURGET1) that best fitted the optical photometry (\cite{rufener89}) for the star. The second was obtained using the same method, but substituting the Kurucz synthetic spectrum by the mean of two low resolution spectra of $\alpha$ Cir calibrated in flux (\cite{aleks96}; \cite{burnashev85}). The two values obtained for the $B.C.$ were $0.15 \pm 0.02$ and $0.23 \pm 0.02$, respectively. 

To place $\alpha$ Cir in the H.R. diagram, the results obtained for the bolometric fluxes were combined with the Hipparcos parallaxe. For the effective temperature we adopted the value from \cite{Kupka96}, $T_{\rm eff} = 7900 \pm 200$ K.
\begin{figure}
\centerline{\includegraphics[width=7cm]{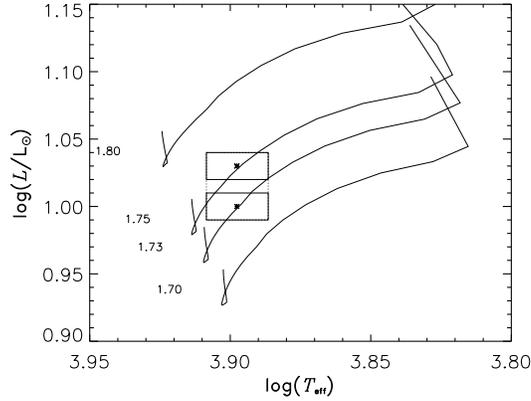}}
\caption{The position of $\alpha$ Cir in the HR diagram with four evolution tracks for models with masses of 1.70, 1.73, 1.75 and 1.80 M$_\odot$ for comparison. The two error boxes correspond to two values of the luminosity (and associated error) derived from two different bolometric fluxes (see text for details).}
\label{fig_HR}
\end{figure}

Both values derived for the $B.C.$ have uncertainties that go beyond their formal errors. In the first case, uncertainties are likely associated with the use of Kurucz models which are not apropriate for peculiar stars. In the second case, additional uncertainties are also expected because the errors associated with the calibration in flux of the low resolution spectra were not provided in the catalogues. Hence, we took a conservative approach and considered both values when placing the star in the HR diagram.

\acknowledgements
MSC is supported by the EC's FP6, FCT and FEDER (POCI2010) through the HELAS international collaboration and through the projects POCI/CTE-AST/57610/2004 and PTDC/CTE-AST/66181/2006 FCT-Portugal.

\end{document}